\documentclass[aps,pre,twocolumn,amssymb,amsfonts,floatfix,showpacs,superscriptaddress]{revtex4}

\usepackage{amsmath} 
\usepackage{amsthm} 
\usepackage{amssymb}	
\usepackage{graphicx} 
\usepackage{dcolumn}
\usepackage{times}
\usepackage{epstopdf}

\usepackage[dvips]{color}

\begin{document}

\title{Effects of the interplay between initial state and Hamiltonian on the thermalization of isolated quantum many-body systems}

\author{E. J. Torres-Herrera}
\author{Lea F. Santos}
\affiliation{Department of Physics, Yeshiva University, New York, New York 10016, USA}

\begin{abstract}
We explore the role of the initial state on the onset of thermalization in isolated quantum many-body systems after a quench. The initial state is an eigenstate of an initial Hamiltonian $\hat{H}_I$ and it evolves according to a different final Hamiltonian $\hat{H}_F$. If the initial state has a chaotic structure with respect to $\hat{H}_F$, i.e., if it fills the energy shell ergodically, thermalization is certain to occur. This happens when $\hat{H}_I$ is a full random matrix, because its states projected onto $\hat{H}_F$ are fully delocalized. The results for the observables then agree with those obtained with thermal states at infinite temperature. However, finite real systems with few-body interactions,  as the ones considered here, are deprived of fully extended eigenstates, even when described by a nonintegrable Hamiltonian.  We examine how the initial state delocalizes as it gets closer to the middle of the spectrum of $\hat{H}_F$, causing the observables to approach thermal averages, be the models integrable or chaotic. Our numerical studies are based on initial states with energies that cover the entire lower half of the spectrum of one-dimensional Heisenberg spin-1/2 systems.
\end{abstract}

\pacs{05.30.-d, 05.45.Mt, 05.70.Ln,75.10.Jm, 02.30.Ik}
\maketitle

\section{Introduction}
\label{Sec:intro}

Experiments in optical lattices have enhanced the interest in the nonequilibrium dynamics of isolated quantum systems
\cite{bloch_dalibard_review_08,Cazalilla2011}. New life has been brought, for example, to the old pursuit of deriving thermodynamics from first principles. One of the main questions in this context refers to the conditions for an isolated quantum system initially far from equilibrium to reach thermal equilibrium.  How decisive are regimes, initial states, and observables? What is universal? 

A general picture has been developed associating chaos with the onset of thermalization~\cite{Deutsch1991,Srednicki1994,zele,ZelevinskyRep1996,Flambaum1996,Flambaum1997,BGIC98,I01,Rigol2008}, but several examples have been promoted as violations of this picture. Cases have been presented of chaotic Hamiltonians where thermalization was not observed for certain initial states~\cite{Roux2010,Biroli2010,Gogolin2011,Banuls2011,Brandino2012} and of integrable systems where thermalization appeared to be reached~\cite{Cassidy2011,Fitzpatrick2011,Rigol2012,Deng2011,He2012,He2013}.

It is important to emphasize that the mere appearance of signatures of chaos associated with the bulk of eigenvalues, such as level repulsion or rigidity~\cite{Guhr1998}, is not sufficient for the onset of thermalization for any initial state. Real systems have few-body and usually short-range interactions. Contrary to full random matrices, their density of states has a Gaussian shape~\cite{French1970,Brody1981} and their delocalized eigenstates are restricted to a region close to the middle of the spectrum~\cite{ZelevinskyRep1996,Flambaum1996,Kota2001,Santos2010PRE,Santos2010PREb}. If the initial state has energy close to the edges of the spectrum, the main contributions to its evolution may come from localized eigenstates, and thermalization may not occur. 

In full random matrices, the eigenstates are completely extended pseudo-random vectors. In real chaotic systems, even in the middle of the spectrum, the eigenstates are limited by the energy shell and their components may show correlations~\cite{Flambaum1993}. Yet, if the initial state is close to the middle of the spectrum, thermalization is expected to happen. Thermalization requires a substantial sampling of delocalized eigenstates, but the latter do not need to be completely extended. It is then natural to presume that thermalization could happen also for integrable Hamiltonians with delocalized eigenstates, provided the initial state be nearly chaotic. 

The reasoning above shows a change in perspective with respect to the condition for thermalization, putting emphasis also on the initial state rather than on the regime of the Hamiltonian only. In the case of an integrable system, it was shown that the quasi-particle number operator behaves thermally for a thermal (mixed) initial state at sufficiently high temperature~\cite{Deng2011}. Subsequently, it was argued that this would hold only for infinite temperature~\cite{He2012}. For a pure initial state corresponding to an eigenstate of an arbitrary Hamiltonian, it was supported that thermalization could happen if this Hamiltonian was chaotic~\cite{Rigol2012,He2013,Khatami2013}, but not if it was integrable~\cite{He2013}. Notice, however, that the initial states considered in \cite{He2013} were not very close to the  middle of the spectrum.

As we stress in this work, an important factor for thermalization is the structure of the initial state with respect to the Hamiltonian that evolves it. Thermalization  is assured for any initial state that fills the energy shell ergodically. In this case its coefficients become uncorrelated random variables following a Gaussian distribution~\cite{Casati1993,Casati1996,Santos2012PRL,Santos2012PRE}. Any eigenstate from a full random matrix projected on the energy eigenbasis of a delocalized system falls in this category. In fact, such states are similar to thermal states at infinite temperature~\cite{Zangara2012,Zangara2013}. The main question is then how close to such scenario of random vectors does an initial state need to be for the viability of thermalization.

We explore this idea in the context of quenches. The system is initially in an eigenstate of an initial Hamiltonian. A parameter is instantaneously changed, which leads to a different final Hamiltonian and starts the system evolution. Four quenches are considered: from an integrable to a chaotic Hamiltonian and vice versa, and from an integrable interacting Hamiltonian (nonmappable to a noninteracting model) to an integrable noninteracting Hamiltonian and vice versa. We analyze how the structure of the initial state changes as it approaches the middle of the spectrum of the final Hamiltonian. In finite systems with few-body interactions, this state will never be completely extended, not even when one of the Hamiltonians involved in the quench is chaotic. We study how close to the center of the spectrum one needs to go for nearly chaotic initial states to emerge and if any of the quenches precludes their existence. 

The system considered is a one-dimensional lattice of spins-1/2. For each quench, we examine initial states with energies ranging from very close to the ground state to very close to the middle of the spectrum. As the initial state approaches the center of the spectrum, the results for the observables get less dependent on position and momentum. Independently of the Hamiltonian, the results move toward those from thermal averages, although not so clearly for all observables. Whether this picture will hold in the thermodynamic limit requires scaling analysis, which is currently unfeasible for systems with interactions, where only small system sizes are available.

This paper is organized as follows. Section \ref{Sec:therm} discusses the conditions for thermalization in isolated quantum systems. In Sec.~\ref{Sec:shell}, we introduce the concept of energy shell as employed here and how to quantify the level of delocalization of the initial state. Section~\ref{Sec:model} describes the model
and quenches investigated. Section\ref{Sec:numerics} shows the numerical results and finishes with a general discussion.  Concluding remarks are presented in Sec.~\ref{Sec:summary}.

\section{Thermalization after a quench}
\label{Sec:therm}

The unitary time evolution of an isolated quantum system out of equilibrium is given by
\begin{equation}\label{eq:instate}
|\Psi(t)\rangle=e^{-i\hat{H}_\text{F}t}|\Psi(0)\rangle=\sum_{\alpha} C_{\alpha}^{\text{ini}}  e^{-iE_\alpha t}|\psi_\alpha\rangle .
\end{equation} 
Above and hereafter, $\hbar$ is set to 1. The initial state  $|\Psi(0)\rangle = |\text{ini}\rangle$ is as an eigenstate of an initial Hamiltonian $\hat{H}_\text{I}$. The dynamics is caused by the sudden change (quench) of some parameter(s) of this Hamiltonian in a time interval much shorter than any characteristic time scale of the model. This results in the final Hamiltonian $\hat{H}_\text{F}$ with eigenvalues $E_\alpha$ and eigenstates $|\psi_\alpha\rangle\neq| \text{ini} \rangle$. The coefficients $C_{\alpha}^{\text{ini}} = \langle \psi_{\alpha} | \text{ini} \rangle $  are the overlaps of the initial state with the eigenstates of $\hat{H}_\text{F}$. 

The dynamics of the expectation value of an observable $\hat{O}$ is computed as
\begin{equation}\label{eq:barwq}
\begin{split}
\langle\hat{O}(t)\rangle
&=\langle\Psi(t)|\hat{O}|\Psi(t)\rangle\\
&=\sum_{\alpha} |C_{\alpha}^{\text{ini}}|^2 O_{\alpha \alpha} + \sum_{\alpha \neq \beta} C_{\alpha}^{\text{ini*}} C_{\beta}^{\text{ini}} e^{i(E_\alpha-E_\beta)t} O_{\alpha \beta}\;,
\end{split}
\end{equation}
where $O_{\alpha \beta} = \langle\psi_\alpha|\hat{O}|\psi_\beta \rangle$ corresponds to the matrix elements of $\hat{O}$ in the energy eigenbasis. In the absence of too many degeneracies, $E_\alpha\neq E_\beta$ for most states, the off-diagonal elements of $\langle\hat{O}(t)\rangle$ oscillate very fast and cancel out on average, so the infinite time average is given by
\begin{align}\label{eq:diagonal}
   \overline{O} &= \lim_{t\rightarrow \infty} \frac{1}{t}
\int^{t}_0 d\tau\,  \langle\hat{O}(\tau)\rangle \nonumber\\
&=\sum_\alpha |C_{\alpha}^{\text{ini}}|^2 O_{\alpha\alpha} \equiv \langle \hat O \rangle_{\rm DE}.
\end{align}
Since the steady state average depends only on the diagonal elements of $\hat{O}$, it is commonly referred to as the prediction from the diagonal ensemble (DE) \cite{Rigol2008}.

Thermalization implies that the expectation value of the observable is close to the thermal average most of the time. This holds under two essential conditions:

(1) The fluctuations of $\langle\hat{O}(t)\rangle$ about the infinite time average are small and vanish in the thermodynamic limit. We can then talk about equilibration in a probabilistic sense. The rate of decay of these fluctuations with system size depends on the system investigated~\cite{Srednicki1999,Reimann2008,Short2011,Short2012,Gramsch2012,Venuti2013,HeSantos2013,Zangara2013}. The important fact for us here is that for the models we study the temporal fluctuations are indeed small and should vanish for very large systems~\cite{Zangara2013}.

(2) The infinite time average and the thermal average are very close in finite systems and eventually coincide in the thermodynamic limit, 
\begin{equation}\label{eq:therm}
\langle\hat{O}\rangle_{\text{DE}}=\langle\hat{O}\rangle_{\text{ME}}.
\end{equation}
Above,
\begin{equation}\label{eq:micro}
\langle\hat{O}\rangle_\text{ME}  \equiv  \frac{1}{{\cal{N}}_{E^{\text{ini}},\Delta  E}}\hspace{-0.5cm}\sum_{\substack{\alpha \\ |E^{\text{ini}}-E_\alpha|<\Delta E}}\hspace{-0.5cm} O_{\alpha\alpha}
\end{equation}
is the average obtained with the microcanonical ensemble, which is the appropriate ensemble when dealing with isolated systems. $E^{\text{ini}}$ is the mean energy of the initial state and ${\cal{N}}_{E^{\text{ini}},\Delta E}$ stands for the number of energy eigenstates in the window $\Delta E$. This window is sufficiently small compared with the energy spectrum, but large enough to contain many energy states. 

The DE depends on the initial state via the components $|C_{\alpha}^{\text{ini}}|^2$, while the ME depends only on the energy. For initial states narrow in energy~\cite{noteSuppl}, two scenarios may lead to the proximity between the two ensembles:

(1) The eigenstate expectation value of the observable, $O_{\alpha \alpha}$, is a smooth function of energy. When the values of $O_{\alpha \alpha}$ do not fluctuate for states close in energy, the result from a single eigenstate inside the microcanonical window  agrees with the microcanonical average. This is the point of view of the eigenstate thermalization hypothesis (ETH) \cite{Deutsch1991,Srednicki1994,Rigol2008,rigol09STATa,rigol09STATb,Yukalov2011}, which certainly holds when the eigenstates of the system are random vectors. Computing $O_{\alpha \alpha}$ with one or another random state leads to very similar results. However, for realistic chaotic systems, one needs to be careful with the region of the spectrum that is considered. As mentioned before, nearly chaotic eigenstates are not ubiquitous to chaotic systems with few-body interactions. They are found away from the edges of the spectrum and the exact region of their predominance depends on the strength of the interaction. 

(2) The coefficients $C_{\alpha}^{\text{ini}}$'s behave as random variables. This happens when the initial state fills the shell. The fluctuations of the coefficients become uncorrelated with possible fluctuations of $O_{\alpha \alpha}$. In Eq.~(\ref{eq:diagonal}), we can then deal with the average of $|C_{\alpha}^{\text{ini}}|^2$ \cite{Flambaum1997,FlambaumARXIV,Kota2011}. 

If the energy shell is filled and  ETH is satisfied through a chaotic $\hat{H}_F$, it is hard to identify which of the two conditions is the main condition for thermalization. However, if $\hat{H}_F$ is integrable, then it is certainly the filling of the shell [condition (2)] that allows for thermalization.


\section{Energy Shell}
\label{Sec:shell}

The notion of energy shell is often used in quantum chaos. The usual procedure in the field is to separate the total Hamiltonian in an unperturbed part, which describes the noninteracting particles (or quasi-particles), and a perturbation, which represents the inter-(quasi-)particle interactions and may drive the system into the chaotic domain. The Hamiltonian matrix is written in the basis corresponding to the unperturbed vectors, which constitute the mean-field basis. 

A mean-field basis vector $|n\rangle$ projected on the eigenstates $|\psi_{\alpha} \rangle$ of the total Hamiltonian is written as
\[
|n \rangle= \sum_{\alpha} C_{\alpha}^n |\psi_{\alpha}  \rangle.
\]
The smoothened distribution of the components $|C_{\alpha}^n|^2$ in the eigenvalues $E_{\alpha}$ is known as the strength function or local density of states~\cite{Flambaum2000}. The energy shell corresponds to the maximal strength function obtained in the limit of a very strong perturbation, that is when the mean-field part can be neglected in comparison with the perturbed part. The energy shell has a Gaussian shape in the center of the spectrum and delimits the maximum possible spreading of the mean-field basis vectors in $E_{\alpha}$ , as well as the maximum level of delocalization of  $|\psi_{\alpha} \rangle$ in the energies of the mean-field basis vectors. Away from the middle of the spectrum, the distribution of $|C_{\alpha}^n|^2$ in $E_{\alpha}$ becomes  skewed to the left (right) when $|n \rangle$ has low (high) energy~\cite{Flambaum1994}.

The level of delocalization of the states increases with the perturbation, but they do not become totally extended as the eigenstates of full random matrices. When dealing with realistic systems of few-body interactions, a completely chaotic state is defined as a state that fills the energy shell~\cite{Casati1993,Casati1996,Santos2012PRL,Santos2012PRE}. 

We bring the concept of chaotic states in connection with the energy shell to our studies of quantum many-body systems after a quench. The unperturbed part of the Hamiltonian now coincides with $\hat{H}_I$, and the total Hamiltonian is $\hat{H}_F$. The latter is written in the basis corresponding to the eigenstates of the initial Hamiltonian. The maximum possible spreading of an initial state with energy in the middle of the spectrum is given by a Gaussian centered at
\begin{equation}\label{eq:E0}
E^{\text{ini}}=\langle \text{ini} |\hat{H}_\text{F}| \text{ini} \rangle
=\sum_\alpha |C_{\alpha}^{\text{ini}}|^2E_\alpha
\end{equation}
and of variance
\begin{equation}\label{eq:sigma}
\begin{split}
\sigma^2&=\sum_\alpha |C_{\alpha}^{\text{ini}}|^2\left(E_\alpha-E^{\text{ini}}\right)^2 \\
&=\sum_n \langle \text{ini} |\hat{H}_\text{F} |n\rangle \langle n|\hat{H}_\text{F}|\text{ini}\rangle-(E^{\text{ini}})^2\\
&=\sum_{n\neq \text{ini}}|H_{\text{ini},n}|^2.
\end{split}
\end{equation}
The variance of the energy shell corresponds to the energy distribution of the initial state. It depends only on the sum of the square of the off-diagonal elements of the final Hamiltonian~\cite{Flambaum1997}. The values of $E^{\text{ini}}$ and $\sigma$ are straightforward to obtain when $\hat{H}_I$ is already in a diagonal form or trivially solved~\cite{Zangara2013}.

\subsection{Delocalization of the initial state and thermal properties}

To quantify the level of delocalization of the initial state we use the inverse participation ratio,
\begin{equation}
\text{IPR}^{\text{ini}} =\frac{1}{\sum_{\alpha} | C_{\alpha}^{\text{ini}}|^4}.
\label{eq:IPR}
\end{equation}
If the initial state is an eigenstate of a full random matrix, it is maximally delocalized. If the matrix comes from a Gaussian orthogonal ensemble (GOE) \cite{Guhr1998}, its projection on the eigenstates of the systems we study here is again a GOE-type of vector, with probability amplitudes from a normal distribution. In this case, $\text{IPR}^{\text{ini}} \sim {\cal D}/3$ \cite{Izrailev1990}, where ${\cal D}$ is the dimension of the Hamiltonian matrix. If the eigenstate comes from a Gaussian unitary ensemble (GUE), the projection is again a GUE-vector and, as we verified, $\text{IPR}^{\text{ini}} \sim {\cal D}/2$. Both states fill the energy shell ergodically. 

It has been shown that a state corresponding to a random superposition of Ising states manifests thermal features. Specifically, the results for local observables led to the same outcomes obtained with a mixed state at infinite temperature~\cite{Alvarez2008,Alvarez2010,Zangara2012}. The state is constructed so that the probability amplitudes of the basis vectors have all the same absolute value, $1/\sqrt{{\cal D}}$, but differ by a random phase $e^{\mathrm{i}2\pi\varphi}$, where $ \varphi $ is a uniformly distributed random variable in $ [0,1) $. We have checked that the projection of such a state onto the eigenstates of our spin-1/2 Hamiltonians  leads to vectors very similar to GUE eigenstates, with $\text{IPR}^{\text{ini}} \sim {\cal D}/2$. 

We therefore infer that an initial state with $C_{\alpha}^{\text{ini}}$ randomly distributed, as the coefficients of eigenstates from full random matrices, have properties similar to those of thermal states at infinite temperature. This is expected, since on average $|C_{\alpha}^{\text{ini}}|^2$ equals $1/{\cal D}$. We have indeed confirmed that, for the observables studied here, when the initial state is an eigenstate from a GOE,  
\begin{equation}
\langle \hat{O} \rangle_{\text{DE}} \sim \langle \hat{O} \rangle_{T \rightarrow \infty}, 
\label{eq:thermal}
\end{equation}
where $T \rightarrow \infty$ indicates the result for a thermal state $\rho$ at infinite temperature, $\rho = \sum_{\alpha} |\psi_{\alpha}\rangle \langle \psi_{\alpha}|/{\cal D}$.

The initial states considered in this paper come from Hamiltonians which, as those from GOE's, are real and symmetric. But since our $\hat{H}_I$ and $\hat{H}_F$ have only two-body interactions, none of our initial states can reach maximum delocalization with  $\text{IPR}^{\text{ini}} \sim {\cal D}/3$. This does not discard, however, the possibility of having infinite time averages approaching microcanonical averages.

\section{Model and quenches}
\label{Sec:model}

We consider a one-dimensional system with open boundary conditions composed of $L$ coupled spins-1/2  and described by the Hamiltonian,
\begin{equation}\label{eq:Ham}
\hat{H} = \hat{H}_z + \hat{H}_{\text{NN}}+ \lambda \hat{H}_{\text{NNN}},
\end{equation}
where
\begin{eqnarray}
&&\hat{H}_z =  \epsilon \hat{S}_{1}^z
\nonumber\\
&&\hat{H}_{\text{NN}} = J\sum_{i=1}^{L-1} \left( 
\hat{S}_i^x \hat{S}_{i+1}^x + \hat{S}_i^y \hat{S}_{i+1}^y +
\Delta \hat{S}_i^z \hat{S}_{i+1}^z \right),  
\nonumber \\
&&\hat{H}_{\text{NNN}} = J\sum_{i=1}^{L-2} \left(\hat{S}_i^x \hat{S}_{i+2}^x + \hat{S}_i^y \hat{S}_{i+2}^y
+ \Delta \hat{S}_i^z \hat{S}_{i+2}^z  \right). \nonumber
\end{eqnarray}
Above, $\hat{S}^{x,y,z}_i $ are spin operators acting on site $i$. $\hat{H}_z$ indicates the presence of a defect on the first site of the chain. It is generated by applying a magnetic field in the $z$-direction, which is slightly larger than the magnetic field on the other sites. The coupling between nearest neighbors (NN) is given by $\hat{H}_{\text{NN}}$, which contains the flip-flop term $\hat{S}_i^x \hat{S}_{i+1}^x + \hat{S}_i^y \hat{S}_{i+1}^y$  and the Ising interaction $\hat{S}_i^z \hat{S}_{i+1}^z $. $\hat{H}_{\text{NNN}}$ describes the coupling between next-nearest neighbors (NNN).  $\Delta$ is the anisotropy parameter and $\lambda$ refers to the ratio between NNN and NN exchange, and both are assumed positive.
The exchange coupling constant $J$ determines the energy scale and is set to 1. 

When both parameters  $\Delta$ and $\lambda$ are simultaneously equal to zero, $\hat{H}$ describes the XX model and can be mapped onto a system of noninteracting spinless fermions, being trivially solvable~\cite{Jordan1928}. When ${\lambda=0}$ and $\Delta\neq1$, we have the XXZ model, which is still integrable but now solved  by means of the Bethe ansatz~\cite{Bethe1931}. When $\lambda\neq0$ the system is nonintegrable and may become chaotic~\cite{Gubin2012,noteFine}. Notice that we refer here to the XX and the XXZ Hamiltonian when only NN couplings are present, so both cases correspond to integrable models.

The purpose of the small defect on the first site of the chain is to break trivial symmetries, such as parity, conservation of total spin, and spin reversal~\cite{Santos2011,Joel2013}, without breaking the integrability of the system~\cite{Alcaraz1987}. We fix $\epsilon=0.1J$. Conservation of total spin in the $z$ direction, $\hat{{\cal{S}}}^z=\sum_n\hat{S}_i^z$, is a remaining symmetry. We deal with the subspace that has $L/3$ up spins, implying dimension ${\cal D}=L!/[(2L/3)!(L/3)!]$ and ${\cal{S}}^z=-L/6$. 

Four quenches are investigated.  Two of them involve only integrable Hamiltonians $(\lambda =0)$, $\Delta$ being the parameter changed:
\begin{itemize}
\item[(a)]  $\Delta_\text{I}= 0 \rightarrow \Delta_\text{F}= 1.5$: from the XX model (noninteracting Hamiltonian) to the XXZ model (interacting Hamiltonian).
\end{itemize}
\begin{itemize}
\item[(b)] $\Delta_\text{I}= 1.5 \rightarrow \Delta_\text{F}= 0$: from the XXZ model (interacting Hamiltonian) to the XX model (noninteracting Hamiltonian).
\end{itemize}
The other two quenches have fixed the value of the anisotropy, $\Delta=0.5$, so that the Hamiltonians are always interacting ones. The quenches involve one strongly chaotic Hamiltonian $(\lambda =1)$ and an integrable one:
\begin{itemize}
\item[(c)]  $\lambda_\text{I}=1\rightarrow \lambda_\text{F}=0$: from a chaotic Hamiltonian to the integrable XXZ model.
\end{itemize}
\begin{itemize}
\item[(d)] $\lambda_\text{I}=0\rightarrow \lambda_\text{F}=1$: from the XXZ Hamiltonian to a chaotic system.
\end{itemize}

Notice that for the quenches above, the perturbation is strong, as shown in Refs.~\cite{Santos2012PRL,Santos2012PRE}. As the perturbation increases from zero, the shape of the initial state (that is, the distribution of the components $|C_{\alpha}^n|^2$ in $E_{\alpha}$)  expands from a delta function to a Breit-Wigner (Lorentzian) form and eventually becomes a Gaussian function as determined by the energy shell. In this paper, we are already in the limit where a Gaussian shape is expected.

\section{Numerical results}
\label{Sec:numerics}

We analyze a broad range of initial states with energies covering the lower half of the spectrum of the final Hamiltonian. They correspond to eigenstates of $\hat{H}_I$ and are selected according to the value of their energy with respect to $\hat{H}_F$, from very close to the ground state to very close to the middle of the spectrum.  For each temperature $T \in [0.1,100]$ (Boltzmann constant $k_{\text{B}}=1$), we select the state whose energy $E^{\text{ini}}$ is closest to
\begin{equation}
\label{eq:energies}
E = \frac{1}{Z} \sum_\alpha E_\alpha e^{-E_\alpha/T},
\end{equation}
where  $Z=\sum_\alpha e^{-E_\alpha/T}$ is the partition function. Low temperatures mean energies close to the edge of the spectrum, while high temperatures correspond to energies close to the middle of the spectrum.

We start by studying how the structure of the initial states with respect to $\hat{H}_F$ changes as $T$ increases and then compare it with the results for the observables. To have a quantitative estimate of the proximity between the results from the diagonal ensemble and those from the microcanonical ensemble, we compute their relative difference. For a generic observable $\hat{O}$, this difference is defined by
\begin{equation}\label{eq:reldif}
\delta\ O=\frac{|\langle\hat O\rangle_{\text{DE}}-\langle\hat O\rangle_{\text{ME}}|}{|\langle\hat O\rangle_{\text{DE}}|}\:.
\end{equation}
To compute the microcanonical averages, we used $\Delta E=0.4$. We verified that the results were not dependent on this value.

Unless indicated, the figures are obtained for $L=18$ leading to ${\cal D} =18\, 564$. Full exact diagonalization is used.

\subsection{Structure of the initial state}

\begin{figure}[htb]
\centering
\vskip 0.5 cm
\includegraphics[width=0.45\textwidth]{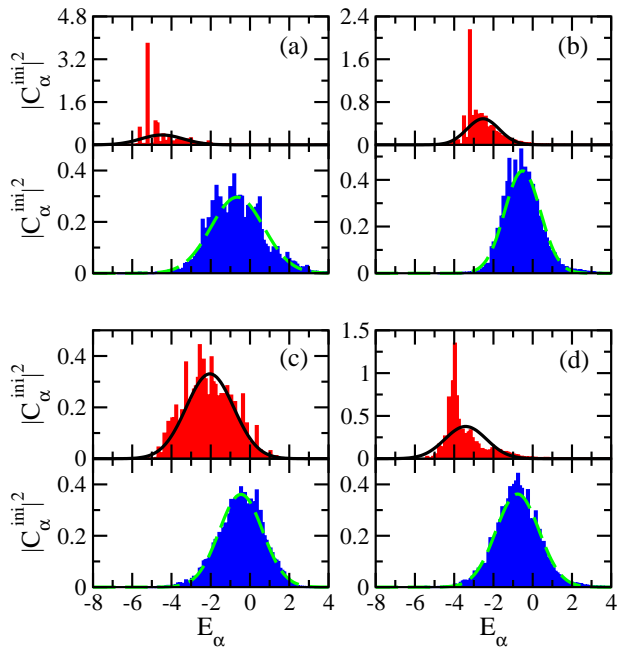}
\caption{(Color online) Distribution of the components of the initial state in the energy representation compared with the energy shell. The pairs of top and bottom panels are labeled according to the quenches, as in Sec.~\ref{Sec:model}:  (a) XX $\rightarrow $ XXZ; (b) XXZ $\rightarrow $ XX; (c) chaotic $\hat{H}$ $\rightarrow $ XXZ; (d) XXZ $\rightarrow $ chaotic $\hat{H}$. Results for two different temperatures are shown: $T = 0.7$  in the top panel of the pair (energy shell in black solid line) and $T = 4.0$ in the bottom panel (energy shell in green dashed line). The energies and variances for low and high temperature are, respectively, :  (a) $E^{\text{ini}}\approx -4.48$, $\sigma^2\approx1.17$ and $E^{\text{ini}}\approx -0.64$, $\sigma^2\approx1.81$. (b) $E^{\text{ini}}\approx -2.53$, $\sigma^2\approx 0.67$ and $E^{\text{ini}}\approx -0.52$, $\sigma^2\approx 0.83$. (c) $E^{\text{ini}}\approx -2.04$, $\sigma^2\approx 1.45$ and $E^{\text{ini}}\approx -0.46$, $\sigma^2\approx1.22$. (d) $E^{\text{ini}}\approx -3.41$, $\sigma^2\approx1.12$ and $E^{\text{ini}}\approx -0.77$, $\sigma^2\approx1.21$.}
\label{fig:shell}	
\end{figure}

In Fig.~\ref{fig:shell}, we compare the distribution of  $|C_\alpha^{\text{ini}}|^2$'s as function of the eigenvalues $E_\alpha$ with the energy shell for two temperatures and four quenches. In correspondence with the Gaussian density of states~\cite{noteDOS}, the filling of the shell improves as $T$ increases and the state approaches the middle of the spectrum. In the case of quenches involving two integrable Hamiltonians the improvement is substantial [Figs.~\ref{fig:shell} (a) and \ref{fig:shell} (b)]. It is significant also for the quench from an integrable to a chaotic Hamiltonian [Fig.~\ref{fig:shell} (d)], confirming the importance of being away from the edges of the spectrum even when the final Hamiltonian is nonintegrable. For these three quenches, (a), (b) and (d), the skewed shape of the initial state towards low energies is evident when $T$ is small. In contrast, the shape of the initial state coming from a chaotic Hamiltonian is not much affected by the temperature [Fig.~\ref{fig:shell} (c)].

\begin{figure}[htb]
\centering
\includegraphics[width=0.35\textwidth]{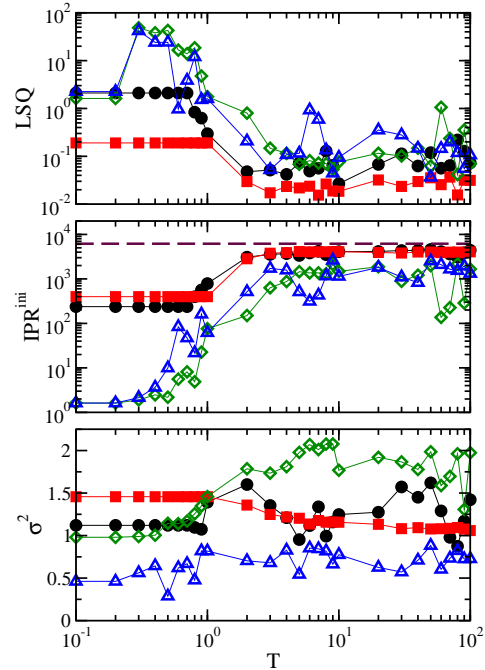}
\caption{(Color online) Least square between the distribution of $|C_{\alpha}^{\text{ini}}|^2$'s and the energy shell (top panel), inverse participation ratio  (middle panel), and variance of the energy shell  (bottom panel) vs $T$ for the following quenches: XX $\rightarrow $ XXZ (green empty diamonds); XXZ $\rightarrow $ XX (blue empty triangles); (c) chaotic $\hat{H}$ $\rightarrow $ XXZ (red filled squares); (d) XXZ $\rightarrow $ chaotic $\hat{H}$ (black filled circles).}
\label{fig:IPR}	
\end{figure}

For a more quantitative picture of the filling of the shell, we show in the top panel of Fig.~\ref{fig:IPR} the least square between the distribution of $|C_{\alpha}^{\text{ini}}|^2$ and the energy shell. The improvement in the filling of the shell with $T$ is evident for all quenches, but the least square values are overall smaller for the quenches involving a chaotic Hamiltonian and more fluctuating for the quenches between integrable Hamiltonians, especially for XXZ $\rightarrow $ XX. We notice that the constant values of the least squares at very low temperatures [more pronounced for quenches (c) and (d)] is due to the lack of states in that region of the spectrum, so close temperatures lead to the selection of the same initial state.

The middle panel of Fig.~\ref{fig:IPR} shows the value of $\text{IPR}^{\text{ini}}$ vs $T$ for the four quenches. It increases with temperature and then saturates. The largest values are reached for the quenches involving a chaotic Hamiltonian, but even in these cases, when $L=18$, $\text{IPR}^{\text{ini}}$ is $\sim 2/3$ of the GOE value ${\cal D}/3$ (the latter is indicated with a thick dashed line in the figure). For the quenches with two integrable Hamiltonians, $\text{IPR}^{\text{ini}}$ is even smaller and large fluctuations are seen, especially for XXZ $\rightarrow $ XX. Therefore, none of the initial states correspond to thermal states at infinite temperature. But more important than the actual value is how $\text{IPR}^{\text{ini}}$ scales with system size. 

The scaling of $\text{IPR}^{\text{ini}}$ with $L$, or equivalently the scaling of the diagonal entropy~\cite{noteShannon} with $L$ \cite{He2013} followed by a comparison with the thermodynamic result, is essential to determining whether thermalization will indeed occur at $L \rightarrow \infty$. If $\text{IPR}^{\text{ini}} \propto {\cal D}$, we would have similar and unbiased participation of the energy eigenstates inside the energy shell and therefore good agreement with microcanonical averages. This would imply also the improving filling of the energy shell with system size, since the width of the shell for the quenches considered must grow linearly with $L$.

\begin{figure}[htb]
\centering
\vskip 0.5 cm
\includegraphics[width=0.45\textwidth]{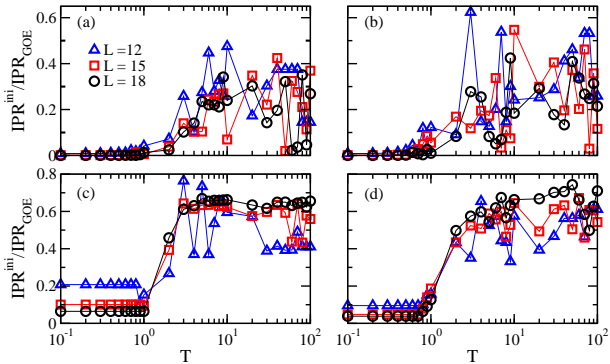}
\caption{(Color online) Ratio between $\text{IPR}^{\text{ini}}$ and $\text{IPR}_{\text{GOE}}$ vs $T$ for three system sizes indicated in panel (a). Quenches as in Fig.~\ref{fig:shell}.}
\label{fig:size}	
\end{figure}

It is an open question whether $\text{IPR}^{\text{ini}} \propto {\cal D}$ for all four quenches. Scaling analyses are not possible when $\hat{H}_F$ is nonmappable to a noninteracting Hamiltonian, since exact diagonalization is required. When comparing the ratio $\text{IPR}^{\text{ini}}/\text{IPR}_{\text{GOE}}$ vs $T$ for $L=12,15,$ and 18 in Fig.~\ref{fig:size}, the latter size leads to an overall smoother behavior and possibly larger values for the quenches involving a chaotic Hamiltonian, whereas large fluctuations remain for the integrable quenches for all system sizes, preventing any general statement. 

\begin{figure}[htb]
\centering
\vskip 0.5 cm
\includegraphics[width=0.45\textwidth]{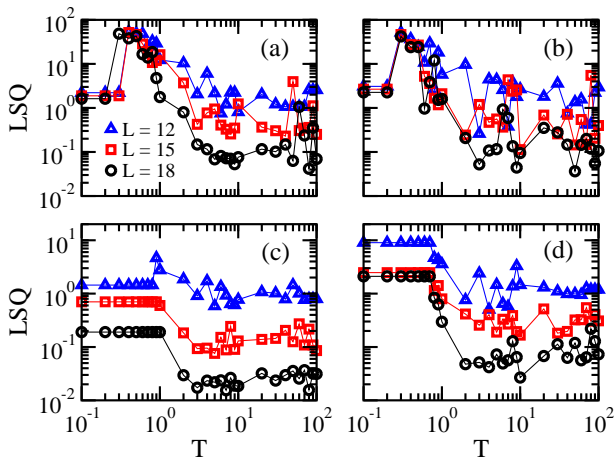}
\caption{(Color online) Least square between the distribution of $|C_{\alpha}^{\text{ini}}|^2$'s and the energy shell vs $T$ for three system sizes indicated in panel (a). Quenches as in Fig.~\ref{fig:shell}.}
\label{fig:LSQvsL}	
\end{figure}

On the other hand, the filling of the energy shell indeed improves with $L$ for all four quenches. This is shown in Fig.~\ref{fig:LSQvsL}, where three system sizes are considered. The improvement is significant for quenches involving a chaotic Hamiltonian and also for the XX $\rightarrow$ XXZ case. The usual large fluctuations of the XXZ $\rightarrow$ XX quench are present, but even here an improvement is noticeable. If the initial state is truly chaotic in the sense of filling the energy shell, an unbiased sampling of the energy eigenstates should occur and lead to thermalization. The results of Fig.~\ref{fig:LSQvsL} point in this direction.

It is informative to look also at the width $\sigma $ of the energy shell as a function of temperature (bottom panel of Fig.~\ref{fig:IPR}). According to Eq.~(\ref{eq:sigma}), the width of the shell is related to the number of states that are directly coupled with the initial state, that is, it quantifies the level of connectivity of the initial state~\cite{Santos2012PRE,Santos2012PRL}. The lowest connectivity is seen for the quench XXZ $\rightarrow $ XX, which also fluctuates significantly. In contrast, $\sigma^2 $ for the reverse quench, XX $\rightarrow $ XXZ, reaches the largest values.  The smoothest behavior with temperature occurs for the quench where $\lambda_I = 1$, which leads also to the best filling of the shell and lowest values of least squares.

From the above results, one cannot rule out the possibility of thermalization for the four quenches considered, provided the initial state is away from the edges of the spectrum. One can, however, suspect the quenches involving only two integrable Hamiltonians, especially the case XXZ $\rightarrow $ XX. For this quench, not only do the three quantities in Fig.~\ref{fig:IPR} fluctuate significantly, but scaling analysis in Ref.~\cite{He2013} discourages expectations for thermalization. In comparison, the quantities in Fig.~\ref{fig:IPR} for the reverse case, XX $\rightarrow $ XXZ, fluctuate less in the experimentally accessible region of temperatures in $1\leq T \leq 10$. This is probably related to the fact that, despite being both integrable, the XX and the XXZ model are very different, the first having a much higher number of degeneracies~\cite{Zangara2013}.

\subsection{Local magnetization} 
\begin{figure}[htb]
\includegraphics[width=0.45\textwidth]{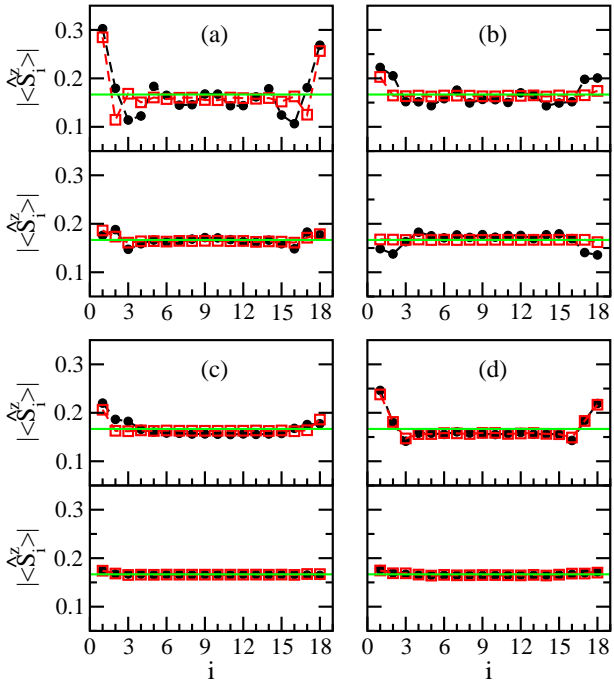}
\caption{(Color online.) Absolute value of the local magnetization for all sites. Comparison between the profiles resulting from the diagonal (black circles) and microcanonical (red squares) ensembles. Green solid line is the homogeneous result -1/6. Quenches as in Fig.~\ref{fig:shell}. For each quench, the upper panel shows results for $T=0.7$ and the lower panel shows results for $T=4.0$.}
\label{fig:mag}
\end{figure}

The local magnetization, $\hat{S}_i^z$, of each site $i$ is shown in Fig.~\ref{fig:mag}  for the diagonal (circles) and microcanonical (squares) ensembles. For each quench, (a), (b), (c), and (d), the upper panel depicts results for an energy close to the edge of the spectrum ($T=0.7$), while the lower panel is for an energy close to the middle of the spectrum ($T=4.0$).
\vskip 0.3 cm

When the temperature is low, the local magnetization shows a strong dependence on the site. This happens for all quenches and is more pronounced at the borders of the chain. As the energy of the initial state approaches the middle of the spectrum, the dependence on the sites almost disappears. A slightly larger value remains on site 1, because of the inclusion of the small defect on the chain [Eq.~\eqref{eq:Ham}]. For all quenches at $T=4.0$, the results for the diagonal ensemble become close to those from the microcanonical ensemble. However, the results are smoother for quenches involving a chaotic Hamiltonian than for those where both Hamiltonians are integrable.  

\begin{figure}[htb]
\includegraphics[width=0.45\textwidth]{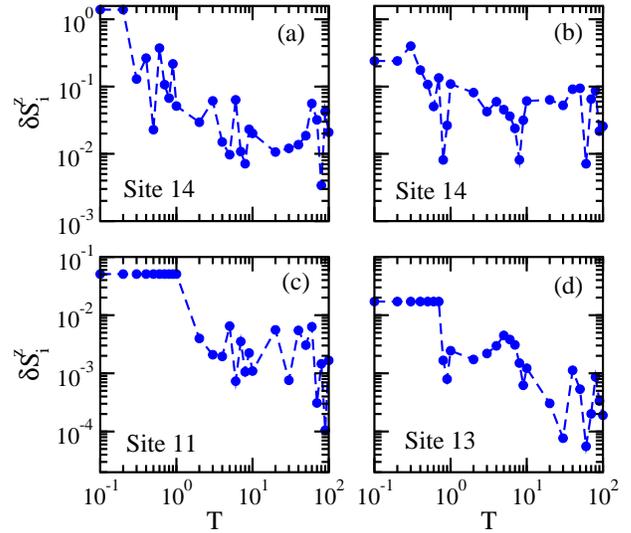}
\caption{(Color online.) Relative difference of the local magnetization vs $T$.  Quenches as in Fig.~\ref{fig:shell}. The site considered is indicated in the panels.}
\label{fig:DMag}
\end{figure}

The local magnetization is a good observable for illustrating finite size effects. In the thermodynamic limit, it is irrelevant if the boundaries are periodic or open, all sites should have the same value,
\[
\langle \hat{S}^z_i \rangle = -1/6,
\]
which is indicated in the figure with a solid line.
The fact that we are further from this scenario for quenches (a) and (b) shows how sensitive quenches with only integrable Hamiltonians are to finite system effects.
The relative difference [Eq.~(\ref{eq:reldif})] for the local magnetization, shown in Fig.~\ref{fig:DMag},  indicates that this problem is more evident for quench (b) [XXZ $\rightarrow $ XX] .
For each quench in the figure, we choose a site for which the decay of $\delta S_i^z$ with temperature is noticeable. For quenches (a), (c) and (d), despite fluctuations, various sites show similar behavior, the infinite time average approaching the thermal result (and the homogeneous -1/6 value) as $T$ increases. For quench (b) the decay for most sites is not visible due to large fluctuations. The site selected is one of the best behaved. Notice also that, when comparing quenches between integrable Hamiltonians with quenches with one chaotic Hamiltonian, $\delta S_i^z$ for the latter is approximately one order of magnitude smaller.
\vskip 0.35 cm
 
\subsection{Spin-spin correlation} 
\begin{figure}[htb]
\includegraphics[width=0.45\textwidth]{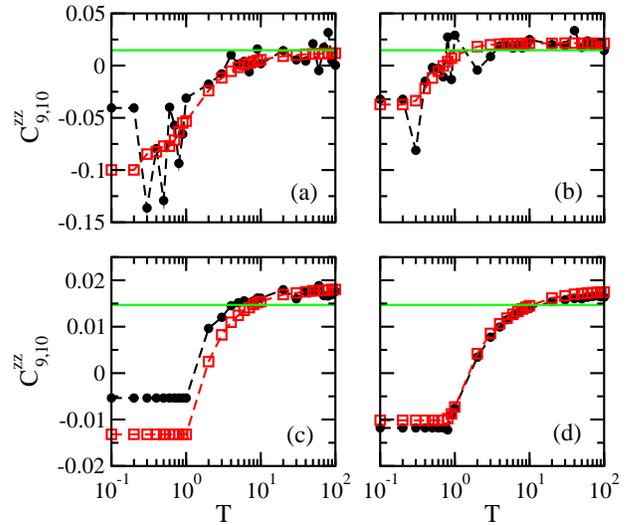}
\caption{(Color online.) Longitudinal spin-spin correlation $C^{zz}_{L/2,L/2+1}$ vs $T$ for the diagonal (black circles) and microcanonical (red squares) ensembles. Quenches as in Fig.~\ref{fig:shell}. The green solid line indicates the result for a thermal state at infinite temperature.}
\label{fig:Czz}
\end{figure}

The spin-spin correlations between neighboring sites in the middle of the chain are given by 
\begin{equation}\label{cxx} 
C^{\mu\mu}_{L/2,L/2+1}= \langle \hat{S}^\mu_{L/2} \hat{S}^\mu_{L/2+1} \rangle,\quad \mu =x,z.
\end{equation}
Figure~\ref{fig:Czz} shows the behavior of the longitudinal spin-spin correlation, $C^{zz}_{L/2,L/2+1}$, vs temperature. As $T$ increases, $C^{zz}_{L/2,L/2+1}$ approaches the result for an initial thermal state at infinite temperature, 
\[
( C^{zz} _{L/2,L/2+1} )_{T\rightarrow \infty} = \frac{L-9}{36 (L-1)},
\]
but does not equal it, indicating that the initial states considered are not completely extended.
The value for $(C^{zz} _{L/2,L/2+1} )_{T\rightarrow \infty}$ is obtained assuming that the average of $|C_{\alpha}^{\text{ini}}|^2$ is $1/{\cal  D}$, as in the case of eigenstates from full random matrices \cite{ZelevinskyRep1996}  [see discussions in Sec.~III.A)]. 
Comparing the results from quenches with two integrable Hamiltonians [Figs.~\ref{fig:Czz} (a) and \ref{fig:Czz} (b)] with those with one chaotic $\hat{H}$ [Figs.~\ref{fig:Czz} (c) and \ref{fig:Czz} (d)], we see that the latter are smoother, especially for the case $\lambda_I =0 \rightarrow \lambda_F=1$. For this quench, the results from DE for the correlations between second and third neighbors are also smoother than for the other quenches and very similar to the results from the ME (not shown).

\begin{figure}[htb]
\vskip 0.2cm
\includegraphics[width=0.45\textwidth]{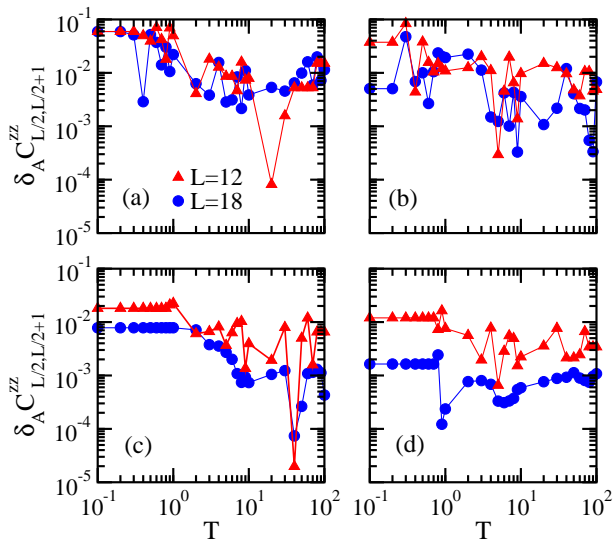}
\caption{Absolute diference between $(C^{zz}_{L/2,L/2+1})_{\text{DE}}$ and $(C^{zz}_{L/2,L/2+1})_{\text{ME}}$  vs $T$ for two system sizes indicated in panel (a). Quenches as in Fig.~\ref{fig:shell}.}
\label{fig:DCzz}
\end{figure}

In Fig.~\ref{fig:DCzz} we show the absolute value of the difference,
\begin{equation}
\delta_A C_{L/2,L/2+1}^{zz}=\left|(C_{L/2,L/2+1}^{zz})_\text{DE}- (C^{zz}_{L/2,L/2+1})_\text{ME}\right|
\end{equation}
versus temperature. Since correlations can go to zero, this is a more appropriate quantity than the relative difference. The results for $\delta_A C_{L/2,L/2+1}^{zz}$ mirror the filling of the energy shell. As the initial state approaches the middle of the spectrum, the difference between diagonal and microcanonical averages, despite fluctuating, decreases for all quenches. The values are smaller for the quenches with $\lambda \neq 0$ [Figs.~\ref{fig:DCzz} (c) and \ref{fig:DCzz} (d)]. In the particular case of XXZ $\rightarrow $ chaotic $\hat{H}_{F}$ [Fig.~\ref{fig:DCzz} (d)],  the decay with $T$ is not significant, being slightly perceptible in the region $1\leq T \leq10$, but comparing with the other quenches, it reaches the smallest values of $\delta_A C_{L/2,L/2+1}^{zz}$.

Figure~\ref{fig:DCzz} shows results for two system sizes, $L=12$ (triangles) and $L=18$ (circles). The decay of $\delta_A C_{L/2,L/2+1}^{zz}$ with $L$ is evident for the quenches involving a chaotic Hamiltonian [Figs.~\ref{fig:DCzz}  (c) and (d)], especially for XXZ $\rightarrow $ chaotic $\hat{H}_{F}$ [Figs.~\ref{fig:DCzz} (d)], therefore indicating the coincidence of DE and ME in thermodynamic limit. For the integrable quenches, the decay with system size is far from obvious, but still conceivable.

As for the transverse spin-spin correlation, it vanishes at high temperature, but the behavior of $\delta_A C_{L/2,L/2+1}^{xx}$ with temperature is similar to that of $\delta_A C_{L/2,L/2+1}^{zz}$.

\subsection{Structure factor} 

\begin{figure}[h]
\includegraphics*[width=0.45\textwidth]{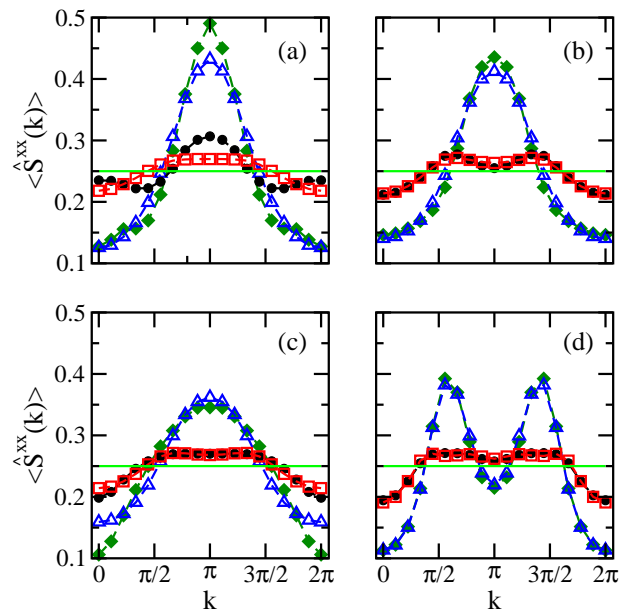}
\caption{(Color online.) Transverse component of the structure factor, $S^{xx}(k)$. Quenches as in Fig.~\ref{fig:shell}. Two temperatures are shown. Filled symbols are for the diagonal ensemble: $T=0.7$ (green diamonds) and  $T=4.0$ (black circles). Empty symbols are for the microcanonical ensemble: $T=0.7$  (blue triangles) and $T=4.0$ (red squares). The green solid line indicates the result for a thermal state at infinite temperature.}
\label{fig:Stxx}
\end{figure}

The structure factor, defined as the Fourier transform of the spin-spin correlations \cite{PhysRevB.24.1429}, corresponds to
\begin{equation}\label{eq:structure}
\hat{S}^{\mu\mu}(k)=\frac{1}{L}\sum_{l,j=1}^L \hat{S}_l^\mu \hat{S}_j^\mu e^{-ik\left(l-j\right)},\quad \mu =x,z.
\end{equation}
Figure~\ref{fig:Stxx} shows results for the diagonal (full symbols) and microcanonical (empty symbols) ensembles for the transverse structure factor $\hat{S}^{xx}(k)$ and two temperatures. The results for $\hat{S}^{zz}(k)$ (not shown) are qualitatively similar, apart from the value for momenta $k=0,2\pi$, which is constant and given by $\langle \hat{S}^{zz}(0,2\pi) \rangle =L/36$ for ${\cal S}^z=-L/6$. 

As seen in the figure, when the energy of the initial state is close to the edge of the spectrum ($T=0.7$), both $\langle \hat{S}^{xx}(k)\rangle_{\text{DE}}$ and $\langle \hat{S}^{xx}(k)\rangle_{\text{ME}}$  have a strong dependence on $k$ for all quenches. This dependence was observed also in Ref.~\cite{canovi_rossini_11}, where the initial state was prepared in the ground state of $\hat{H}_I$. When $\hat{H}_F$ has only nearest neighbor couplings, a single peak emerges for $k=\pi$, whereas two peaks exist when second neighbors are included. As $E^{\text{ini}}$ approaches the middle of the spectrum (illustrated for $T=4.0$), the $k$-dependence decreases drastically for all quenches.  The results get close, but not equal, to that for a thermal state at infinite temperature, $\langle \hat{S}^{xx}(k) \rangle_{T\rightarrow \infty} =1/4$.

\begin{figure}[htb]
\vskip 0.3 cm
\includegraphics[width=0.45\textwidth]{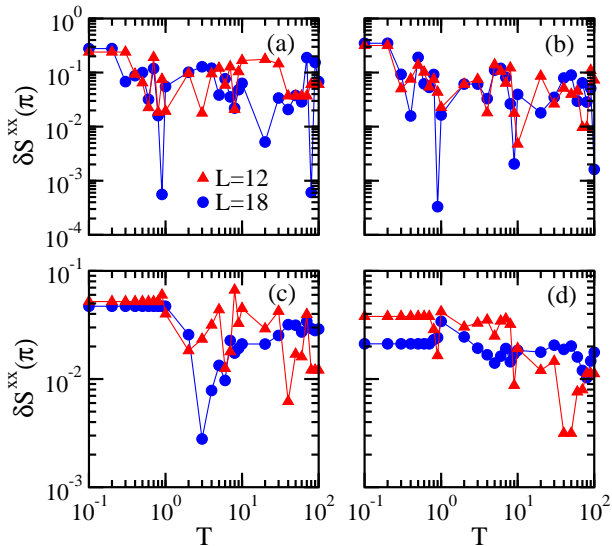}
\caption{(Color online.) Relative difference for the structure factor in the $x$ direction vs $T$ for two system sizes indicated in panel (a). Quenches as in Fig.~\ref{fig:shell}.}
\label{fig:DStxx}
\end{figure}

\begin{figure}[htb]
\vskip 0.3 cm
\includegraphics[width=0.45\textwidth]{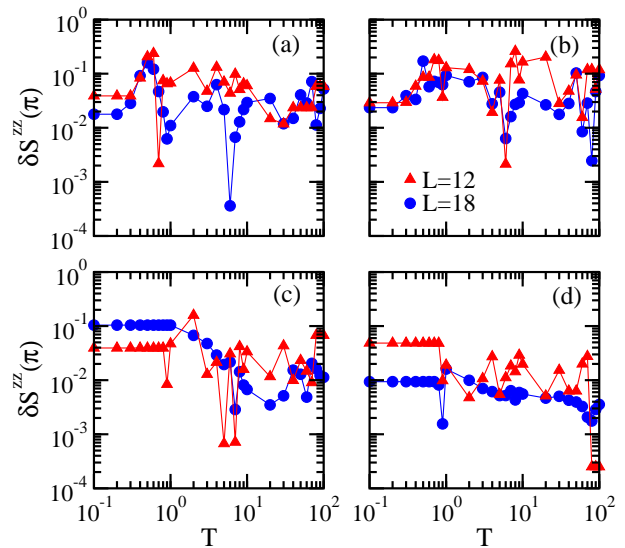}
\caption{(Color online.) Relative difference for the structure factor in the $z$ direction vs $T$ for two system sizes indicated in panel (b). Quenches as in Fig.~\ref{fig:shell}.}
\label{fig:DStzz}
\end{figure}

In Figs.~\ref{fig:DStxx} and~\ref{fig:DStzz}, we show $\delta S^{xx}(\pi)$ and $\delta S^{zz}(\pi)$, respectively. The decay with temperature for these observables is overall less pronounced than for the spin-spin correlation. It is particularly subtle for the integrable quenches. In this case, it is questionable that one can indeed discern a decay amid the fluctuations. 

The values of $\delta S^{\mu \mu }(\pi)$ reached for the quenches involving $\lambda \neq 0$ [panels (c) and (d) in Figs.~\ref{fig:DStxx} and \ref{fig:DStzz}] are, once again, smaller than for quenches between integrable Hamiltonians [(a) and (b)]. In comparing the directions, the values of $\delta S^{z z }(\pi)$ [Figs.~\ref{fig:DStzz} (c) and \ref{fig:DStzz} (d)] are smaller than those for $\delta S^{xx }(\pi)$ [Figs.~\ref{fig:DStxx} (c) and \ref{fig:DStxx} (d)]. 

With respect to system size, the decay with $L$ is noticeable in panels (c) and (d) in Figs.~\ref{fig:DStxx} and \ref{fig:DStzz}. For the integrable quenches, it is also suggested for $S^{z z }(\pi)$ for XX $\rightarrow $ XXZ [Fig.~\ref{fig:DStzz} (a)]. Nevertheless, even if a decay with $L$ really exists for the integrable quenches in some regions of the spectrum, it must be much slower then when a chaotic Hamiltonian is present.


\subsection{Discussion} 

As the temperature increases, the results for the observables after relaxation become smoother in position and momentum. They approach the results obtained with thermal states at infinite temperature, but do not equal them. Even though the initial states for all quenches get more delocalized as they get closer to the middle of the spectrum, they are not equivalent to the totally delocalized eigenstates from full random matrices. 

For spin-spin correlations, the difference between diagonal and microcanonical ensemble decreases with $T$ for all quenches, but for the structure factor, the picture is less clear. $\langle \hat{S}^{\mu \mu }(\pi) \rangle_{\text{DE}}$ approaches $\langle \hat{S}^{\mu \mu }(\pi) \rangle_{\text{ME}}$, especially in the $z$ direction, for quenches involving one chaotic Hamiltonian, whereas for quenches between two integrable Hamiltonians,  large fluctuations prevent conclusive statements. 

Our results give strong indications that the differences between the infinite time average and the thermodynamic ensemble decrease with system size when one of the Hamiltonians in the quench is chaotic. For the integrable quenches, the fluctuations are large, although a decay is suggested in Fig.~\ref{fig:DStzz} (a). The better filling of the energy shell with $L$ for all four quenches also gives support to the possibility of thermalization at $L \rightarrow \infty $ even for integrable quenches. However, detailed scaling analyses are still missing and they are essential for final conclusions.

In Ref.~\cite{He2012}, a study of how the temperature of the initial state affects the properties of the system was also developed. The initial states were thermal (mixed) states evolving according to a Hamiltonian for hard-core bosons mapped onto noninteracting fermions. Scaling analysis is possible in this case. It was shown that thermalization could only happen for thermal states at infinite temperature. The same may or not be the case when interactions are present, such as for the XXZ model. As mentioned, the XX and the XXZ models differ significantly in terms of degeneracies~\cite{Zangara2013}.

We notice that several works have shown that thermalization does not happen for quenches between two noninteracting integrable Hamiltonians~\cite{Cassidy2011,rigol06STATb,Rigol2007,Calabrese2011,Calabrese2012,Caux2013}. In the present work we have studied quenches between different integrable models, including interacting ones~\cite{KormosARXIV,Fagotti2013,Pozsgay2013}, or between an integrable and a chaotic Hamiltonian. These scenarios have been much less explored.

\section{Conclusions}
\label{Sec:summary}

An important factor determining the viability of thermalization in isolated quantum many-body systems is the chaotic structure of the initial state with respect to the Hamiltonian that evolves it. If the initial state fills the energy shell, thermalization may happen even when ETH is not satisfied, as in cases where the final Hamiltonian is integrable. 

Here, we studied the scenario of strong perturbation, where the expected shape of the initial state is Gaussian. This is, however, not any Gaussian function, but the one corresponding to the energy shell with width given by the energy distribution of the initial state. We analyzed how the filling of the energy shell improves as the initial state approaches the middle of the spectrum and how this affects the results for different observables.  None of the finite systems considered, where only few-body interactions are present, lead to completely extended initial states. Thus, none of them correspond to thermal states at infinite temperature. Nevertheless, they further delocalize as the temperature increases for all quenches considered, those involving only integrable Hamiltonians: XX $\rightarrow$ XXZ and XXZ $\rightarrow$ XX, and those with one chaotic Hamiltonian: XXZ $\rightarrow$ chaotic model and chaotic Hamiltonian $\rightarrow$ XXZ.

As the initial state further spreads out with temperature, the results for the observables become smoother and the infinite time averages (diagonal ensemble) approach the thermal averages (microcanonical ensemble). This behavior is more evident when one of the Hamiltonians of the quench is chaotic. In the presence of one chaotic Hamiltonian, the approach of the two ensembles as system size increases is also more perceptible.

Scaling analyses are crucial to resolving whether thermalization can indeed occur for the four quenches studied here. When comparing $L=$12, 15, and 18, we see that the filling of the energy shell improves for all quenches, this being more significant when the final Hamiltonian is interacting. Such spreading of the initial state with respect to the shell suggests that the diagonal ensemble should indeed approach the thermodynamic ensemble as $L$ increases, because filling of the shell implies an unbiased sampling of the energy eigenstates. Nevertheless, larger system sizes and detailed scaling analysis for the observables are necessary to make this statement definitive.

Another essential aspect for future studies is the specification of the threshold in temperature above which thermalization holds and how it depends on the model and system size. If thermalization can occur only for initial states too close to the middle of the spectrum, the temperatures involved are unrealistically large and not of practical use. 
\vskip 0.5 cm
\vskip 0.5 cm

\begin{acknowledgments}
\vskip -0.3 cm
This work was supported by NSF Grant No.~DMR-1147430. We thank Kai He, Felix Izrailev, and Marcos Rigol for useful discussions.
\end{acknowledgments}


\end{document}